\documentclass[prd,showpacs,showkeys,preprint]{revtex4}

\usepackage{amsmath}
\usepackage{amssymb}
\usepackage{yfonts}[1988/10/03]
\usepackage{graphicx}
\newcommand {\beq}{\begin{equation}}
\newcommand {\eeq}{\end{equation}}
\newcommand {\beqa}{\begin{eqnarray}}
\newcommand {\eeqa}{\end{eqnarray}}

\begin{document}
\title{A new experiment for the gravitational waves detection }
\author{ Basem Ghayour$^{1}$\footnote{ba.ghayour@gmail.com}, Jafar Khodagholizadeh$^{2}$\footnote{gholizadeh@ipm.ir}, Christian Corda$^{3}$\footnote{cordac.galilei@gmail.com},  Ming-Lei Tong$^{4}$\footnote{mltong@ntsc.ac.cn}, Ali Ghayour$^{5}$\footnote{ali.ghayour@std.kashanu.ac.ir}}
\affiliation{$^{1}$ School of Physics, University of Hyderabad,
Hyderabad-500 046. India}, 

\affiliation{$^{2}$Farhangian University, P.O. Box 11876-13311, Tehran,
Iran},  

\affiliation{$^{3}$International Institute for Applicable Mathematics \&
Information Sciences (IIAMIS), B.M. Birla Science Centre, Adarsh Nagar,
Hyderabad - 500 463, India and Dipartimento di Matematica e Fisica, Istituto Livi, Via Antonio Marini,
9,59100 Prato, Italy},  

\affiliation{$^{4}$National Time Service Center, Chinese Academy of
Sciences, Xi'an 710600, China.}

\affiliation{$^{5}$University of Kashan, P.O. Box  8731753153, Kashan, Isfahan
Iran}


\date{\today} 

\begin{abstract}
A new experiment for the gravitational waves (GWs) detection is proposed. It is indeed shown that the effect of GWs on sound waves (SWs)
in a fluid is that GWs vary the pressure of the fluid by crossing it. This
variation can be found by analysing the gauge of the local observer. It is
shown that one can, in principle, detect GWs through the proposed new
experiment. The variation of the pressure of the fluid, which represents
detected signals, are indeed much higher than the correspondent
values of GWs amplitudes. The examples of rotating neutron stars (NSs)
and  relic GWs are discussed. Remarkably, a confrontation of the
proposed new method with a previous paper of Singh et al. on a similar
approach shows a possible improvement of the sensitivity concerning the
potential detection of GWs. It must be emphasized that this proposed procedure may be difficult
in practical experiments because of the presence of different types of noise.
For this reason, a Section of the paper is dedicated to the discussion of such noises. On the other hand, this paper must be considered as being
pioneering in the new proposed approach. Thus, we hope in future, more
precise studies of the noise which concerns the proposed new experiment will be done.

\end{abstract}
\maketitle
\section{\label{sec:level1}Introduction }
Gravitational waves (GWs) emissions were indirectly discovered from the compact binary system
PSR1913+16, composed by two neutron  stars (NSs) \cite{1}. Such
a discovery, which permitted to award the Nobel  Prize of Physics in  1993
to Russell Hulse and Joseph Taylor, excited interest in GWs science
despite the first efforts at direct GWs detection started before it
\cite{2}. Those efforts involved the design, implementation,
and advancement of extremely sophisticated GWs detection technology
which is requested by researchers working in this field of research
\cite{2}. The most important reason for GWs research is, to use
GWs as a probe of the systems that produce them. The famous event
GW150914, which is the first observation of GWs from a binary black
hole (BH) merger \cite{3} occurred in the 100th anniversary of
Albert Einstein's prediction of GWs \cite{4}. That event was
a cornerstone for science in general and for gravitational physics
in particular. Indeed it  gave definitive proof of the existence of
GWs, the existence of BHs that having mass greater than 25 solar masses
and the existence of binary systems of BHs which merge in a time less
than the age of the Universe \cite{3}. Such a direct GWs discovery,
represented the starting of the new era of the GWs astronomy and it enabled
 Rainer Weiss, Barry Barish
and Kip Thorne to win the Nobel Prize of Physics in  2017  . After the event GW150914, the LIGO Scientific Collaboration
announced other  new GWs detections \cite{5}.

There are lot of experiments for the direct GWs detection with different
methods. Ground-based laser interferometers like Advanced LIGO \cite{6},
VIRGO \cite{7}, GEO \cite{8}, TAMA\cite{9}, DECIGO
\cite{10}, AIGO \cite{11} and space-based laser interferometers
like LISA \cite{12}, eLISA \cite{13} and Big Bang Observer
\cite{14} which are the most famous. This kind of GWs detectors could
be, in principle decisive to confirm the physical consistence of
the general theory of relativity (GTR) or alternatively to endorse
the framework of extended theories of gravity \cite{15,16}.
In fact, some differences between the GTR and alternative theories
can be pointed out in the linearized theory of gravity through different
interferometer response functions \cite{15,16}. A controversial
issue on a potential GWs  consists in the detection
of the B-modes of the polarization of the cosmic microwave background
 \cite{17}. More precise measures will be needed to confirm
such a GWs in the future \cite{18}. Other attempts
are based on measurements of polarization of electromagnetic waves
\cite{19,20} and on maser beam passing through a strong static
magnetic field \cite{21,22}. But based on the weakness of GWs
amplitudes, researchers prefer using laser interferometer technology \cite{5,6}.  Also there is another method that analysed the
sensitivity to continuous-wave strain fields of a kg-scale optomechanical system
formed by the acoustic motion of super fluid helium-4 parametrically coupled to
a superconducting microwave cavity, see \cite{a1} for more details. But again the sensitivity based on this method was  low.

Therefore in this paper, we attempt to introduce another experiment for a potential
GWs detection. The gravity effect of GWs on longitudinal waves like
sound waves (SWs) in a fluid will be considered. GWs perturb the shape of SWs and
this perturbation can vary the pressure in the fluid. The effect of
this perturbation can be found by solving the geodesic equation. Remarkably,
the estimated amounts of the pressure based on this perturbation are
much higher than the strain sensitivity of GWs interferometers and another methods.  For examples a confrontation of the proposed new method in \cite{a1} on a similar approach shows a possible improvement of the sensitivity
concerning the potential detection of GWs,  The key point is that one
can find superposition in the intersection points between the two types
of wave. Such a superposition causes a variation in the SWs pressure. 
Thus GWs can be in principle, detected by measuring this perturbation.

The paper is organized as it follows. In Sec. 2, SWs will be reviewed.
In Sec. 3, the set up of GWs detection will be analysed by considering
the perturbation on SWs. In Sec. 4, the different sources of noises
will be investigated and in Sec. 5 the conclusion remarks will be
discussed.

\section{Sound waves}\label{two}
Plane waves such as SWs are longitudinal waves that require a material
medium (a fluid) to exist. We start to review the driving of
plane waves in the fluid in brief. Some parameters will be used as
it follows: \medskip{}

$x\equiv$coordinate of the particle in its situation of equilibrium; 

$w\equiv$variation of the particle with respect to $x$;

$u_{x}\equiv\frac{\partial w}{\partial t}\equiv$instant speed of
particle;

$\rho_{0}\equiv$density of fluid in its situation of equilibrium;

$P\equiv$instant pressure of each point of the fluid;

$P_{0}\equiv$pressure of the fluid in it's situation of equilibrium;

$p\equiv dP=P-P_{0}\equiv$variation of the pressure;

$\gamma\equiv$speed of propagation of the wave. \medskip{}

In this paper, we define ``a particle'' as being a small volume
of the fluid if one can assume that there is no variation for the
pressure, density and speed of all molecules of that volume. In addition,
we consider the following assumptions for measuring the variation of pressure: 
\begin{itemize}
\item  for the sake of simplicity, the fluid must be homogeneous, isotropic and elastic;

\item in order to  consider the effect  of GWs  on SWs, the amplitude of SWs must be small;  
\item also for same reason, the variation of the density must be small with respect to the its value
at the equilibrium, see tables. [1,2] in Sec.(3) for some numerical examples of pressure.
\end{itemize}
Then, let us consider a layer of the fluid in equilibrium state with a vertical section
$S$ between two parallel surfaces with positions $x$ and $x+dx$,
see Fig.[1]. 
\begin{figure}\label{figg}
\includegraphics[scale=0.7]{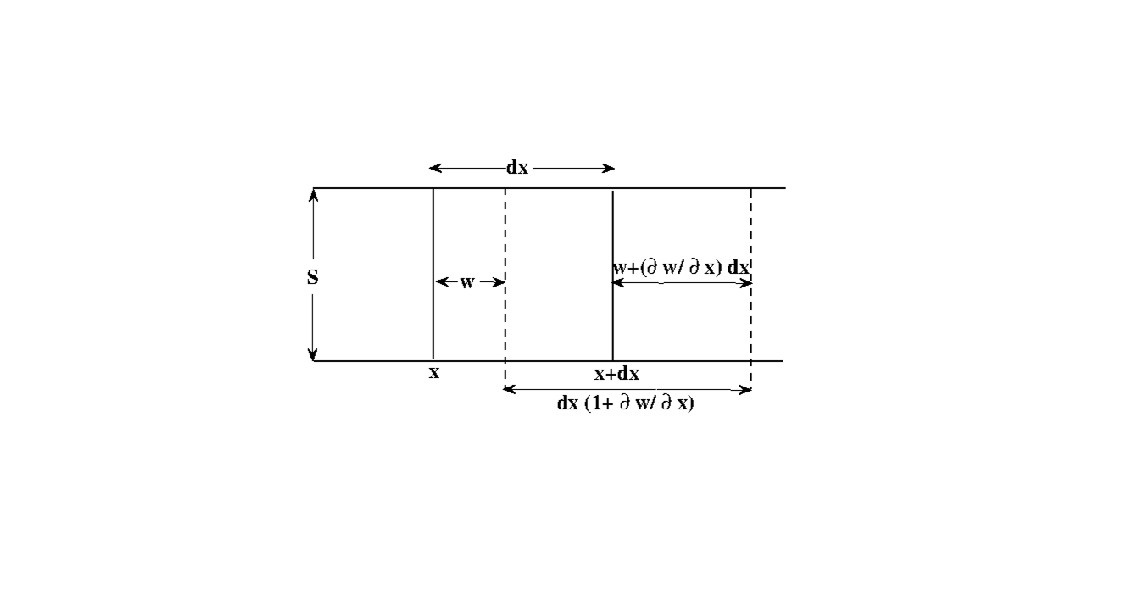}
\caption{The variation of the longitudinal positions of the plane wave in the
fluid.}
\end{figure}
Therefore, the mass of this layer is $\rho_{0}Sdx$. Now, let us suppose
this layer moves to $w, w+\left(\frac{\partial w}{\partial x}\right)dx$
due to right - pass wave. Because the pressures on the both sides of layer are not equal, therefore there exists a force that moves the
mass of the layer $\rho_{0}Sdx$ to the right side as it follows: 

\begin{equation}
dF_{x}=[p-(p+\frac{\partial p}{\partial x }dx)]S=-\frac{\partial p}{\partial x}dx S.\label{eq: 1}
\end{equation}
This force causes an acceleration $\frac{\partial^{2}w}{\partial t^{2}}$.
Then from Newtons second law with $m=\rho_{0} dx S$, one gets
\begin{equation}
-\frac{\partial p}{\partial x}=\rho_{0}\frac{\partial^{2}w}{\partial t^{2}}.\label{eq: 2}
\end{equation}
There is another relation for the pressure (see  Appendix.[A] for more details): 

\begin{equation}
p=-\rho_{0}\gamma^{2}\frac{\partial w}{\partial x}.\label{eq: 3}
\end{equation}
By combining Eqs. (\ref{eq: 2}) and (\ref{eq: 3}) one gets the plan
wave equations: 
\begin{equation}
\frac{\partial^{2}w}{\partial t^{2}}=\gamma^{2}\frac{\partial^{2}w}{\partial x^{2}}\label{eq: 4}
\end{equation}

\begin{equation}
\frac{\partial^{2}p}{\partial t^{2}}=\gamma^{2}\frac{\partial^{2}p}{\partial x^{2}},\label{eq: 5}
\end{equation}
where the speed of propagation of the wave $\gamma$ is given by $\left(\frac{dP}{d\rho}\right)_{0}$
\cite{23}.

\section{The effect of gravitational waves on sound waves}\label{yy}
In order to rigorously derive the effect of GWs on SWs, one needs
to perform the fluid dynamics calculation shown in Section 2 in the
space-time of the GWs. The calculation shown in Section 2 was indeed
performed with the implicit assumption of Minkowskian space-time.
In a weak gravitational field (which is the GWs case), the metric
is almost Minkowskian \cite{24} 
\begin{equation}
g_{\mu\nu}=\eta_{\mu\nu}+h_{\mu\nu},\label{eq: 6}
\end{equation}
where $\eta_{\mu\nu}$ is the standard metric tensor of the flat
Minkowskian space-time (In this space-time, Lorentz transformations can be visualized as ordinary rotations of the four dimensional Euclidean sphere  $x^2+y^2+z^2-c^2t^2=$ constant,  where c is the speed of light)  and  $h_{\mu\nu}\ll1$ is the weak GWs
perturbation \cite{24}. Also the $h_{\mu\nu}$ satisfy the linearized
Einstein equation $\square h_{\mu\nu}=0$ \cite{24}. By assuming
that GWs travel in the $z-$direction, in the TT-gauge the line element
which describes the plane polarized GWs propagating in flat space-time
is given by \cite{24} 
\begin{equation}
h_{\mu\nu}=\left(\begin{array}{cccc}
0 & 0 & 0 & 0\\
0 & h_{+} & h_{\times} & 0\\
0 & h_{\times} & -h_{+} & 0\\
0 & 0 & 0 & 0
\end{array}\right)\exp [i(kz-\omega t)],\;\;\;\;
\omega=kc
,\label{eq: 7}
\end{equation}
where $h_{+}$ and $h_{\times}$ are the two standard GWs polarizations.
For astrophysical sources it should be $h_{+}\thickapprox h_{\times}\leq10^{-21}$
\cite{3}, \cite{25}-\cite{29}.

On the other hand, as GWs detection is performed in a laboratory environment
on Earth, one typically uses the coordinate system in which space-time
is locally flat and the distance between any two points is given simply
by the difference in their coordinates in the sense of Newtonian physics.
This is the so-called gauge of the local observer \cite{15,24}.
In such a gauge the GWs manifest themselves by exerting tidal forces
on the test masses (here we consider ``particles'', i.e. small volumes
of the fluid). By putting a particle in the origin of the coordinate
system, the components of the separation vector are the coordinates
of the second particle. The effect of the GWs is to drive this particle
to have oscillations. Equivalently we can say that there is a gravitational
potential which generates the tidal forces \cite{24} 
\begin{equation}
V(\overrightarrow{r},t)=-\frac{1}{4}\ddot{h}_{+}(t)[x^{2}-y^{2}],\label{eq:potenziale in gauge Lorentziana}
\end{equation}
 and that the motion of the particle is governed by the Newtonian
equation

\begin{equation}
\ddot{\overrightarrow{r}}=-\bigtriangledown V.\label{eq: Newtoniana}
\end{equation}
In geometric terms the connection between Newtonian theory and linearized
general relativity are given by the relation between the gravitational
potential and the time component of the metric \cite{313} 
\begin{equation}
g_{00}=-1-\frac{2V}{c^{2}}.\label{eq: g00}
\end{equation}
The equations of motion for the particle in the gauge of the local
observer are well known \cite{24} 
\begin{equation}
\ddot{x}=\frac{1}{2}\ddot{h}_{+}x\label{eq: accelerazione mareale lungo x}
\end{equation}

\begin{equation}
\ddot{y}=-\frac{1}{2}\ddot{h}_{+}y.\label{eq: accelerazione mareale lungo y}
\end{equation}
Thus, setting 
\begin{equation}
-\frac{\partial p_{1}}{\partial x}\equiv\frac{1}{2}\rho_{0}\ddot{h}_{+}x;\qquad\ddot{x}\equiv\frac{\partial^{2}w_{1}}{\partial t^{2}},\label{eq: setting}
\end{equation}
one gets the total perturbation (GWs plus SWs) as 
\begin{equation}
-\frac{\partial p}{\partial x}-\frac{\partial p_{1}}{\partial x}=\rho_{0}\frac{\partial^{2}w}{\partial t^{2}}+\rho_{0}\frac{\partial^{2}w_{1}}{\partial t^{2}},\label{eq: 12-1}
\end{equation}
where $p_{1}$ and $w_{1}$ are the pressure and the variation
of the position of the particle due to the GWs presence respectively. (We note that $\partial x$ should be replaced by $\partial x\sqrt{1+h_{+}}$
in the GWs case. But as it is $h_{+}\ll1$ , this will not affect
our results. Therefore, we can ignore it.)
Now, setting 
\begin{equation}
p_{h_{+}}\equiv p+p_{1};\qquad w_{h_{+}}\equiv w+w_{1},\label{eq: setting 2}
\end{equation}
Then Eqs. (\ref{eq: 2}), (\ref{eq: 3}), (\ref{eq: 4}), (\ref{eq: 5}),
change as 
\begin{equation}
-\frac{\partial\left(p_{h_{+}}\right)}{\partial x}=\rho_{0}\frac{\partial^{2}\left(w_{h_{+}}\right)}{\partial t^{2}}\label{eq: 15}
\end{equation}
\begin{equation}
p_{h_{+}}=-\rho_{0}\gamma^{2}\frac{\partial\left(w_{h_{+}}\right)}{\partial x}\label{eq: 16}
\end{equation}
\begin{equation}
\frac{\partial^{2}\left(w_{h_{+}}\right)}{\partial t^{2}}=\gamma^{2}\frac{\partial^{2}\left(w_{h_{+}}\right)}{\partial x^{2}}\label{eq: 17}
\end{equation}
\begin{equation}
\frac{\partial^{2}\left(p_{h_{+}}\right)}{\partial t^{2}}=\gamma^{2}\frac{\partial^{2}\left(p_{h_{+}}\right)}{\partial x^{2}},\label{eq: 18}
\end{equation}
 Now, combining Eq. (\ref{eq: accelerazione mareale lungo x}) with
the second definition in Eq. (\ref{eq: setting}), one gets 
\begin{equation}
\frac{\partial^{2}w_{1}}{\partial t^{2}}=\frac{1}{2}\ddot{h}_{+}x=\frac{1}{2}\omega^{2}h_{0+}x\cos\left(kz-\omega t\right),\label{eq: cos}
\end{equation}
where 
\[
\begin{array}{c}
h_{+}\equiv h_{0+}\cos\left(kz-\omega t\right),\\
\\

\end{array}
\]
Thus, one gets the solution for $w_{1}$ as 

\begin{equation}
w_{1}=\frac{1}{2}xh_{0+}\left[-\cos\left(kz-\omega t\right)+\cos kz+\omega t\sin kz\right].\label{eq: w1}
\end{equation}
One gets also 
\begin{equation}
\begin{array}{c}
p_{1}=-\rho_{0}\gamma^{2}\frac{\partial w_{1}}{\partial x}=\\
\\
\frac{1}{2}\rho_{0}\gamma^{2}h_{0+}\left[\cos\left(kz-\omega t\right)-\cos kz-\omega t\sin kz\right].
\end{array}\label{eq: p1}
\end{equation}
From Eq. (\ref{eq: setting 2}) one can see that, in the GWs absence,
that is $h_{0+}=0$, then $p_{h_{+}}=p$, and $w_{h_{+}}=w.$ 
Setting $\cos\left(kz-\omega t\right)=1$ for
the sake of simplicity, one writes 
\begin{equation}
\begin{array}{c}
\Delta p\equiv p_{h_{+}}-p=\\
\\
\frac{1}{2}\rho_{0}\gamma^{2}h_{0+}\left[1-\left(\cos kz+\omega t\sin kz\right)\right].
\end{array}\label{eq: delta p}
\end{equation}
Now, we will see that from this last relation one can, in principle
detect the effect of  GWs. But, in practice the presence of different types of noise
may cause some problems for realizing this kind of GWs detection.
This issue will be discussed in next Section. On the other hand, we
stress that this paper must be considered as being pioneering in the
new proposed approach. Thus, we hope in future, more precise studies
of the noise which concerns the proposed new experiment. 

One also notes that GWs and SWs are two different waves with transversal
and longitudinal properties respectively. Thus, one cannot merely
realize a superposition between the two different types of waves.
Hence, starting from Eq. (\ref{eq: 12-1}), all the equations involving
the total perturbation (GWs plus SWs) are satisfied only for the intersection
points between GWs and SWs, see Fig.[2]. 
In other words, GWs cause perturbation in the SWs shape. The perturbation
appears in the intersection points (the intersection point stands
for the particle) because SWs feel extra acceleration and/or pressure
in those points. Remarkably, the values of $\triangle p$ are much
higher than the correspondent values of GWs amplitudes.
Thus, it seems that the proposed method has an important advantage
with respect to the standard interferometer technology because in
this new approach one needs to measure much higher quantities with
respect to the very small interferometer strains due to GWs.

\begin{figure}\label{fig}
\includegraphics[scale=0.5]{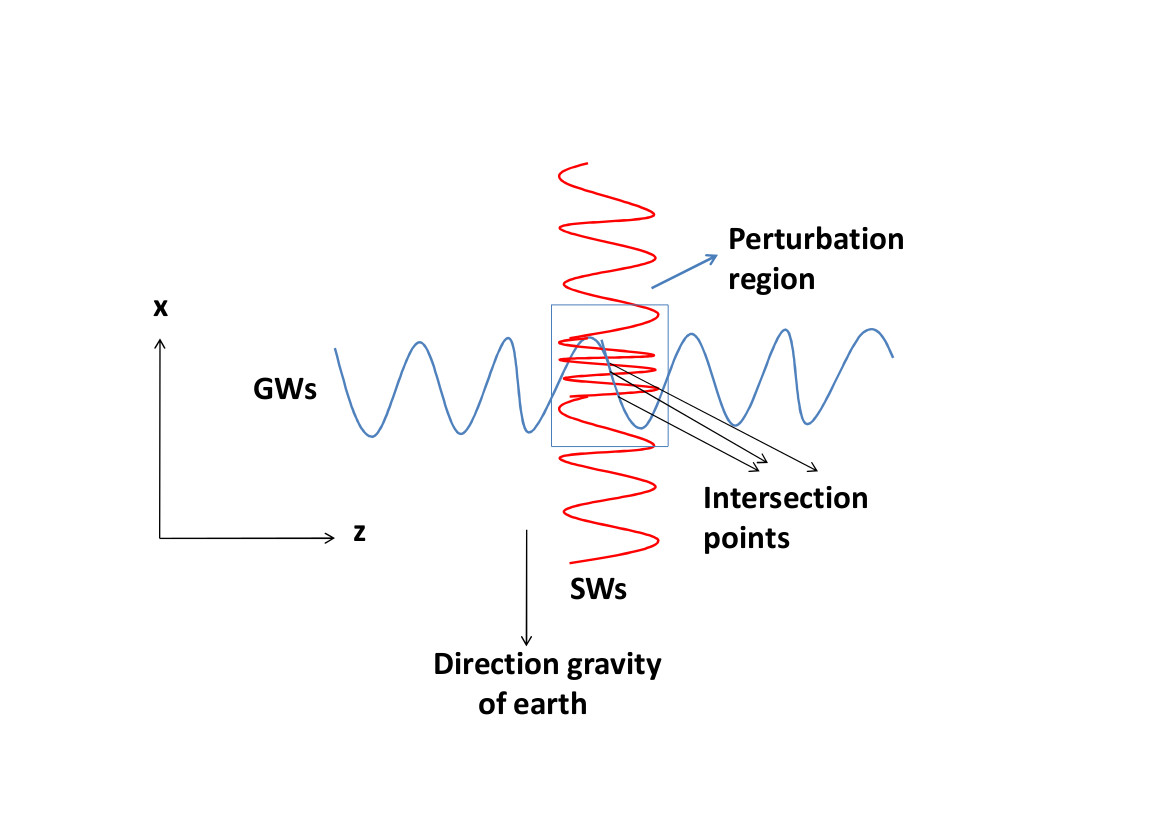}

\caption{Encounter of GWs with SWs. GWs cause perturbation in SWs. This perturbation
appears in the intersection points (the intersection point stands
for the particle). }
\end{figure}

Now, as we propose this new experiment for the GWs detection as being
not alternative but instead, complementary to interferometric GWs
detectors. We use Eq. (\ref{eq: delta p}) to investigate about the
possible detection of GWs from potential astrophysical and cosmological
sources outside the frequency range where interferometers have the
highest sensitivity, which is about $1Hz\leq f\leq100Hz$ \cite{5,6}.
A suitable range is  one of the rotating NSs which till now have
not yet been detected by LIGO. This is in the range of $100Hz\leq f\leq1000Hz$
\cite{5,6}. In that case, any bumps on  imperfections
in the NS spherical shape generate GWs as the NS spins \cite{5}.
If the NS spin rate stays constant, so it emits the  GWs \cite{5}.
Hence GWs are continuously with the same frequency and amplitude. Thus
these are called ``Continuous GWs''\cite{5}. The simplest model
of the NS Continuous GWs emission is given by the so-called rigidly-rotating
aligned triaxial ellipsoid \cite{a2}. The corresponding GW amplitude
depends on the moments of inertia along three principal axes of the
ellipsoid, which characterize the NS ellipticity,  the distance
to the source and  the NS period of rotation \cite{a2}. An estimated upper bound is $h_{0+}\approx10^{-24}$
\cite{a2}. The most rapidly rotating NSs currently
known rotate at order of hundreds Hz \cite{5,6}. Considering
an integration time of order of years and recalling that for the air
it is $\rho_{0}\gamma^{2}\approx10^{5}\,Pa$, while $-1\leq\sin kz\leq1$,
from Eq. (\ref{eq: delta p}) one gets 
\begin{equation}
-10^{-9}\leq\Delta p_{NSs}\leq10^{-9}\,Pa.\label{eq: delta p rivelazione}
\end{equation}
This seems a very small value, but for water we have instead $\rho_{0}\gamma^{2}\approx10^{9}\,Pa$,
which gives 
\begin{equation}
-10^{-5}\leq\Delta p_{NSs}\leq10^{-5}\,Pa.\label{eq: delta p rivelazione acqua}
\end{equation}
The different results for another fluids are write down in Tab.1. The obtained amounts are more higher than corresponds amplitude of mentioned NS above $h_{0+}\approx10^{-24}$. This seems to be an advantage
with respect to standard interferometer technology. 

\begin{table}[t]
{ Table.1. Variation of pressure due to the NSs and relic GWs in different fluids.
Note that the symbol C means centigrade.}\\
{\begin{tabular}{@{}ccccc@{}} \toprule
Matter&$
\rho_{0}\gamma^{2}$&$\;\;\Delta p_{NSs}$&$\Delta p_{relic\; GWs}$\\ \hline
Water&$ 10^9$&\;\;$-10^{-5} \leq\Delta p \leq 10^{-5}$ &  $-10^{-8} \leq\Delta p \leq 10^{-8}$  \\
Air&$10^{5}$&$-10^{-9} \leq\Delta p \leq 10^{-9}$ & $-10^{-12} \leq\Delta p \leq 10^{-12}$ \\
Oil&$ 10^{9}$&$-10^{-5} \leq\Delta p \leq 10^{-5}$ &  $-10^{-8} \leq\Delta p \leq 10^{-8}$  \\
Acetic\; acid&$ 10^{9}$&$-10^{-5} \leq\Delta p \leq 10^{-5}$ &  $-10^{-8} \leq\Delta p \leq 10^{-8}$ \\
Mercury\;&$ 10^{10}$\;&$-10^{-4} \leq\Delta p \leq 10^{-4}$ \;&  $-10^{-7} \leq\Delta p \leq 10^{-7}$ \\
Oxygen (-220\;C)\;&$ 10^{6}$\;&$-10^{-8} \leq\Delta p \leq 10^{-8}$ \;&  $-10^{-11} \leq\Delta p \leq 10^{-11}$ \\
Molten\; lead (340\;C)\;&$ 10^{10}$\;&$-10^{-4} \leq\Delta p \leq 10^{-4}$ \;&  $-10^{-7} \leq\Delta p \leq 10^{-7}$\\
Ether\;&$ 10^{8}$\;&$-10^{-6} \leq\Delta p \leq 10^{-6}$ \;&  $-10^{-9} \leq\Delta p \leq 10^{-9}$ \\
Ethanol\;&$ 10^{9}$\;&$-10^{-5} \leq\Delta p \leq 10^{-5}$ \;&  $-10^{-8} \leq\Delta p \leq 10^{-8}$ \\
Castor\; oil\;&$ 10^{9}$\;&$-10^{-5} \leq\Delta p \leq 10^{-5}$ \;&  $-10^{-8} \leq\Delta p \leq 10^{-8}$\\
\botrule
\end{tabular} \label{sa2}}
\end{table}

The most important cosmological GWs source is given by the relic GWs.
The production of relic GWs is well known in various works in the
literature by using the so called adiabatically-amplified zero-point
fluctuations process, which has been originally developed in the relic
GWs framework by the soviet physicists L. P. Grishchuk \cite{a3}.
Then, it has been shown how the standard inflationary scenario for
the early universe can, in principle provide a distinctive spectrum
of relic GWs \cite{a4}. The potential existence of relic gravitational
radiation arises from general assumptions. It indeed derives from
a mixing between basic principles of classical theories of gravity,
starting from general relativity, and of quantum field theory. The
zero-point quantum oscillations, which produce relic GWs, are generated
by strong variations of the gravitational field in the early universe
\cite{a3,a4}. Then, the detection of relic GWs is the only
way to learn about the evolution of the very early universe, up to
the bounds of the Planck epoch and the initial singularity \cite{a3,a4}.
In more recent years, the analysis has been adapted also to the framework
of extended theories of gravity \cite{a5}. In the standard inflationary
scenario, the relic GWs are characterized by a dimensionless spectrum
\cite{a4,a5}
\begin{equation}
\Omega_{gw}(f)\equiv\frac{1}{\rho_{c}}\frac{d\rho_{gw}}{d\ln f},\label{eq: spettro}
\end{equation}
where 
\begin{equation}
\rho_{c}\equiv\frac{3}{8}H_{0}^{2}\label{eq: densita critica}
\end{equation}
is the critical  energy density of the universe, $H_{0}$ the value
of the Hubble expansion rate and $d\rho_{gw}$ the energy density
of relic GWs in the frequency range $f$ to $f+df$. The spectrum
of relic GWs in inflationary models is flat over a wide range of frequencies,
that is in the range \cite{a4}-\cite{a5} 
\begin{equation}
10^{-16}Hz\leq f\leq10^{10}Hz\label{eq: relic GWs range},
\end{equation}
see fig.[3].

\begin{figure}
\includegraphics[scale=0.8]{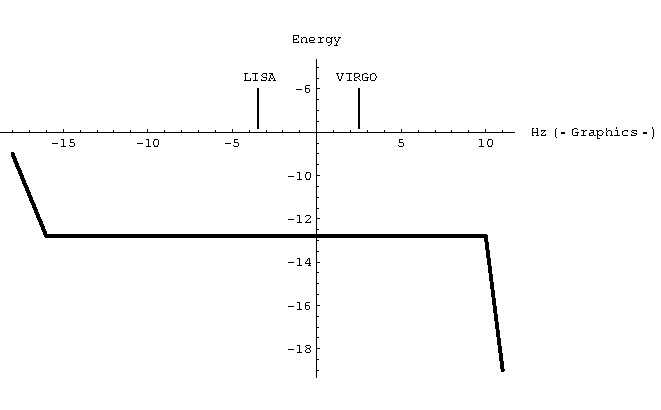}

\caption{Adapted from ref. \cite{a5}.
The spectrum of relic GWs in inflationary models is flat over a wide
range of frequencies. The horizontal axis is $\log_{10}$ of frequency,
in Hz. The vertical axis is $\log_{10}\Omega_{gw}$. The amplitude
of the flat region depends only on the energy density during the inflationary
stage; we have chosen the largest amplitude consistent with the WMAP
constrains on scalar perturbations. This means that in the range $10^{-16}Hz\leq f\leq10^{10}Hz$
it is $\Omega_{gw}(f)h_{100}^{2}\leq9*10^{-13}$. }
\end{figure}
The spectrum is also connected with the characteristic amplitude $h_{C}$
of the relic GWs by the equation \cite{a4,a5} 
\begin{equation}
h_{C}(f)\simeq1.26*10^{-18}\left(\frac{1{\rm Hz}}{f}\right)\sqrt{h_{100}^{2}\Omega_{gw}(f)}.\label{eq: legame ampiezza-spettro}
\end{equation}
A dimensionless factor $h_{100}$ is included. It comes from an uncertainty in the value of $H_{0}$. In fact, for about 50 years in last century,  $H_{0}$ was estimated to be between 50 and 90 (km/s)/Mpc generating a long controversy \cite{h1}. Such a controversy was partially resolved in the late 1990s through very precise cosmological observations due to the development of the Lambda-CDM model. Thus, the value of $h_{100}$ was fixed around 70 (km/s)/Mpc \cite{h2}. On the other hand, a more recent controversy (Hubble tension) started from difference between the estimations on the current value of  $H_{0}$ calculated using CMB ($H_{CMB}= 67.66 \pm 0.42\ \textmd{Km/s/Mpc}$) \cite{cmb}
and SNeIa ($H_{SN}= 74.03 \pm 1.42\ \textmd{Km/s/Mpc}$) \cite{SNeIa}.

In the range (\ref{eq: relic GWs range}) it is \cite{a4,a5}
\begin{equation}
\Omega_{gw}(f)h_{100}^{2}\leq9*10^{-13},\label{eq: Piattezza}
\end{equation}
which permits to write 
\begin{equation}
h_{C}(f)\leq10^{-24}\left(\frac{1{\rm Hz}}{\omega}\right)\label{eq: ampiezza 2}
\end{equation}
in the same range. By combining Eqs. (\ref{eq: delta p}) and (\ref{eq: ampiezza 2})
and considering again an integration time of order of years one gets
for the water 
\begin{equation}
-10^{-8}\leq\Delta p_{relic\; GWs}\leq10^{-8}\,Pa.\label{eq: delta p rivelazione relic GWs}
\end{equation}
On the other hand,
one can consider fluids different from water in order to obtain a
series of results similar to Eqs. (\ref{eq: delta p rivelazione acqua}, \ref{eq: delta p rivelazione relic GWs}).
Such results due to NSs and relic GWs are written down in the Tab.1 for comparison purpose. The more suitable fluids seems to be the Mercury and Molten Lead (340 C). Therefore again our obtained results are higher than the strain based on standard interferometer technology ($h_{0+}\approx10^{-24}$ as a sample) and has an advantage with respect to it.    

\begin{table}[t]
{ Table.2. Variation of pressure due to GWs from different pulsars listed in \cite{a1}
in different fluids.}\\
{\begin{tabular}{@{}ccccc@{}} \toprule
Matter&$
\rho_{0}\gamma^{2}$&$\;\;\Delta p_{pulsar}(J0034-0534)$&\;$\Delta p_{pulsar}(J1301+0833)$\\ \hline
Water&$ 10^9$&\;\;$-10^{-8} \leq\Delta p \leq 10^{-8}$ &  $-10^{-8} \leq\Delta p \leq 10^{-8}$  \\
Air&$10^{5}$&$-10^{-12} \leq\Delta p \leq 10^{-12}$ & $-10^{-12} \leq\Delta p \leq 10^{-12}$ \\
Oil&$ 10^{9}$&$-10^{-8} \leq\Delta p \leq 10^{-8}$ &  $-10^{-8} \leq\Delta p \leq 10^{-8}$  \\
Acetic\; acid&$ 10^{9}$&$-10^{-8} \leq\Delta p \leq 10^{-8}$ &  $-10^{-8} \leq\Delta p \leq 10^{-8}$ \\
Mercury\;&$ 10^{10}$\;&$-10^{-7} \leq\Delta p \leq 10^{-7}$ \;&  $-10^{-7} \leq\Delta p \leq 10^{-7}$ \\
Oxygen (-220\;C)\;&$ 10^{6}$\;&$-10^{-11} \leq\Delta p \leq 10^{-11}$ \;&  $-10^{-11} \leq\Delta p \leq 10^{-11}$ \\
Molten\; lead (340\;C)\;&$ 10^{10}$\;&$-10^{-7} \leq\Delta p \leq 10^{-7}$ \;&  $-10^{-7} \leq\Delta p \leq 10^{-7}$\\
Ether\;&$ 10^{8}$\;&$-10^{-9} \leq\Delta p \leq 10^{-9}$ \;&  $-10^{-9} \leq\Delta p \leq 10^{-9}$ \\
Ethanol\;&$ 10^{9}$\;&$-10^{-8} \leq\Delta p \leq 10^{-8}$ \;&  $-10^{-8} \leq\Delta p \leq 10^{-8}$ \\
Castor\; oil\;&$ 10^{9}$\;&$-10^{-8} \leq\Delta p \leq 10^{-8}$ \;&  $-10^{-8} \leq\Delta p \leq 10^{-8}$\\
\botrule
\end{tabular} \label{ssa2}}
\end{table}

\begin{table}[t]
{ Table.3. Continued Tab.2}\\
{\begin{tabular}{@{}ccccc@{}} \toprule
$\Delta p_{pulsar}(J1747-4036)$&$
\Delta p_{pulsar}(J1902-5105)$&$\;\;\Delta p_{pulsar}(J1939+2134)$&\;$\Delta p_{pulsar}(J1959+2048)$\\ \hline
$-10^{-9} \leq\Delta p \leq 10^{-9}$&$ -10^{-8} \leq\Delta p \leq 10^{-8}$&\;\;$-10^{-8} \leq\Delta p \leq 10^{-8}$ &  $-10^{-8} \leq\Delta p \leq 10^{-8}$  \\
$-10^{-13} \leq\Delta p \leq 10^{-13}$&$-10^{-12} \leq\Delta p \leq 10^{-12}$&$-10^{-12} \leq\Delta p \leq 10^{-12}$ & $-10^{-12} \leq\Delta p \leq 10^{-12}$ \\
$-10^{-9} \leq\Delta p \leq 10^{-9}$&$ -10^{-8} \leq\Delta p \leq 10^{-8}$&$-10^{-8} \leq\Delta p \leq 10^{-8}$ &  $-10^{-8} \leq\Delta p \leq 10^{-8}$  \\
$-10^{-9} \leq\Delta p \leq 10^{-9}$&$ -10^{-8} \leq\Delta p \leq 10^{-8}$&$-10^{-8} \leq\Delta p \leq 10^{-8}$ &  $-10^{-8} \leq\Delta p \leq 10^{-8}$ \\
$-10^{-8} \leq\Delta p \leq 10^{-8}$\;&$ -10^{-7} \leq\Delta p \leq 10^{-7}$\;&$-10^{-7} \leq\Delta p \leq 10^{-7}$ \;&  $-10^{-7} \leq\Delta p \leq 10^{-7}$ \\
$-10^{-12} \leq\Delta p \leq 10^{-12}$\;&$ -10^{-11} \leq\Delta p \leq 10^{-11}$\;&$-10^{-11} \leq\Delta p \leq 10^{-11}$ \;&  $-10^{-11} \leq\Delta p \leq 10^{-11}$ \\
$-10^{-8} \leq\Delta p \leq 10^{-8}$\;&$ -10^{-7} \leq\Delta p \leq 10^{-7}$\;&$-10^{-7} \leq\Delta p \leq 10^{-7}$ \;&  $-10^{-7} \leq\Delta p \leq 10^{-7}$\\
$-10^{-10} \leq\Delta p \leq 10^{-10}$\;&$ -10^{-9} \leq\Delta p \leq 10^{-9}$\;&$-10^{-9} \leq\Delta p \leq 10^{-9}$ \;&  $-10^{-9} \leq\Delta p \leq 10^{-9}$ \\
$-10^{-9} \leq\Delta p \leq 10^{-9}$\;&$ -10^{-8} \leq\Delta p \leq 10^{-8}$\;&$-10^{-8} \leq\Delta p \leq 10^{-8}$ \;&  $-10^{-8} \leq\Delta p \leq 10^{-8}$ \\
$-10^{-9} \leq\Delta p \leq 10^{-9}$\;&$ -10^{-8} \leq\Delta p \leq 10^{-8}$\;&$-10^{-8} \leq\Delta p \leq 10^{-8}$ \;&  $-10^{-8} \leq\Delta p \leq 10^{-8}$\\
\botrule
\end{tabular} \label{sssa2}}
\end{table}

For the sake of completeness, we signal an important previous work \cite{a1} on an
approach similar to the approach in this paper. The Authors of \cite{a1} analysed the
sensitivity to continuous-wave strain fields of a kg-scale optomechanical system
formed by the acoustic motion of super fluid helium-4 parametrically coupled
to a superconducting microwave cavity. Such a narrow-band detection scheme
can operate at very high Q-factors, while the resonant frequency is tunable
through pressurization of the helium in the $0.1-1.5$ kHz range. Consequently,
this kind of detector can, in principle, be tuned to a variety of astrophysical
sources and also remain sensitive to a particular source over a long period of
time. For thermal noise limited sensitivity, strain fields on the order of $h\simeq\frac{10^{-23}}{\sqrt{Hz}}$ could be, remarkably, detectable \cite{a1}. Therefore, the
detector  can compete with interferometric GWs detectors in
particular for certain pulsar sources within a few months of integration time.
Hence in next step and for comparison purpose, we select some of strains of different pulsars that called spin-down from Tab.1 of \cite{a1} to obtain a series of results from Eq.
(\ref{eq: delta p}).
 A confrontation of Tab.1 of \cite{a1} with the our obtained numbers in  Tab.2 shows an improvement of the sensitivity with respect to \cite{a1}. It seems indeed
that the sensitivity arising from the method proposed in this work could be, in
principle, higher than the sensitivity arising from the method proposed in \cite{a1}.
This seems again  to be an advantage
with respect to standard interferometer technology and mentioned above method.



\section{Noise in gravitational waves detectors}

There are multiple sources of noise which can affect the performance
of GWs detections. The more important types of noise in interferometers
are thermal noise, shot noise, seismic noise and radiation pressure
noise. The source of thermal noise arises from three main areas, that
are the pendulum modes of suspension of the mirrors, the internal
modes of the mirrors and the violin modes in the suspension wires.
The shot noise arises from the quantum mechanical fluctuations in
the phase quadrature of the electromagnetic field. The seismic noise
comes from a lack of complete isolation of the mirrors form seismic
activity. Finally, the radiation pressure noise arises from the quantum
mechanical fluctuations in the amplitude quadrature of the electromagnetic
field. 

On the other hand, the detector which is suggested in this paper is
different from the interferometer. The device that we propose, represents
a variant what is referred to as a ``resonant bar"
detector, or``Weber bar", after the scientist
Joseph Weber pioneered their use \cite{a8}, with the important
difference that it uses fluid as the medium instead of a solid such
as aluminium or steel, as has been used traditionally \cite{a8}.
The disadvantage of the fluid will likely have over another medium could
be dissipation. Maybe the fluid will loose energy due to internal
heating at too high a rate to be useful in building up a significant
response to a GW. In fact, the aluminium resonant bar detectors have
relied on the very high $Q$ factor of the material to absorb and
store vibrational energy, building up a large response to a GW having
the correct frequency. Therefore, one must take this into consideration
when computing the sensitivity of the device. On the other hand, a
recent approach in \cite{a9} could in principle permit to solve
this problem. A new concept to drastically reduce acoustic radiation
damping for fluids has been developed in \cite{a9}. A specifically
designed cavity enclosing the resonator seems able to couple the radiated
field back in the resonator \cite{a9}. Experiments on a custom
tuning fork have been carried out and by realizing a cavity tenfold
having a quality factor $Q=75.000$ in air at atmospheric pressure
\cite{a9}. As the order of magnitude of $Q$ factors for aluminium
are $1000<Q<10000$ \cite{a10}, i.e. one order of magnitude less
than the one of the designed cavity in \cite{a10}. One hopes that
the dissipation problem in fluids can be solved in the future.

Another noise that must be taken in account is thermal noise.
In fact, at the scales being discussed in this paper, objects that
we as humans perceive to be solid and stationary are neither \cite{a8}.
At this scale, objects are composed of atoms, each of which is mechanically
excited by thermal energy, and this cannot be ignored \cite{a8}.
Robert Brown indeed detected in 1828 the ``brownian motion"
observing small particles of dust suspended in water \cite{a11}.
Historically, the most remarkable physical interpretation of the ``brownian
motion" arises from a very famous paper of Albert Einstein
\cite{a12}. 

Seismic noise also cannot be ignored. One of the main sources of noise
in this type of detector is indeed the gravity of the Earth. Hence,
the detector must be supported somehow. A proposal in this sense could
be to float it in space.

Thus, it is important to understand what is the order of magnitudes of different noises
in order to see if they are comparable with respect to pressure changes coming from GWs.
In fact, if the pressure changes due to GWs are smaller than those of noises, then using
the experiment proposed in this paper will be impossible. By considering Tables 2 and 3,
one sees that mercury seems to be the most suitable fluid to be used. In fact,for mercury
the order of magnitude of pressure changes coming from GWs is $10^{-7}$. We recall that, in
standard conditions for temperature and pressure, mercury is a very stable liquid largely
used in thermometers, barometers, manometers. In particular and remarkably, in standard
conditions for temperature and pressure, the pressure changes of mercury can be reduced to
the order of magnitude of $10^{-8}$ \cite{49}. This is due to mercury's strong surface tension, high
density and quasi-incompressibility\cite{49}. Thus, it seems that the using of mercury could permit, in principle, to detect GWs via the proposed experiment.

Finally, for the sake of completeness, it is important to stress the following:

Generally, in order to reduce noise:  

1) One could reduce measuring bandwidth and i.e. to measure more slowly. 

2) E.g. In order to obtain low-frequency AC measurements, then the current should
be of low bias. That means that one should use primarily very low
levels of signals. On the contrary, shot-noise is caused by the
fluctuations in the number of photons detected at the photodiode.
Consequently, this noise is minimized when utilizing a large laser
power. Of course, these two cases conclude to diametrically opposed
results relevant to avoiding shot noise, but their common
characteristic is to use outlier conditions in the signal's levels
depending on the type of case. 

3) Also, the only ways to reduce the thermal noise content are to reduce the temperature of operation. 

4) Relevant to confronting the matter of seismic noise one could use the
"Seismic noise-reduction techniques for use with vertical stacking: An
empirical comparison" thus the "Possible earthquake forecasting in a
narrow space-time-magnitude window". 

5) Relevant to "radiation pressure noise", and taking into consideration the Advanced LIGO
interferometers, the quantum radiation pressure noise would be
enhanced in the astrophysically important band of [10, 30] Hz, so the
injection of squeezed states decreasing the shot noise would degrade
the interferometers’  low-frequency sensitivity, see \cite{PT}.

The suggested approaches to reduce noise could be the object of future works.

\section{Discussion and conclusion}
  
In this paper a new experiment for the GWs detection has been proposed.
The proposal is based on the investigation of the perturbation in
a fluid. If GWs cross the fluid, they cause a perturbation in the
shape of SWs in the fluid. This perturbation makes extra pressure
and variations of the position, the velocity and the acceleration
of the particles in the fluid. Thus GWs can be in principle, detected
by carefully measuring these effects. This seems to be an advantage
with respect to standard interferometer technology and another methods such as  \cite{a1},  because one measures
much higher quantities with respect to the very small 
strains due to GWs.

On the other hand, this proposed procedure may be difficult in practical
experiments because of the presence of different types of noise. For
this reason, a section of the paper has been dedicated to the discussion
of such noises. We also stress that this paper must be considered
as being pioneering in the new proposed approach. Thus one hopes that
in future, more precise studies of the noise which concerns the proposed
new experiment will be done. 

\section{Acknowledgements }
The Authors thank the Referees for very useful comments and suggestions.


\appendix \label{m}
 \section {}
 When the surface of wave moves in direction of x axis, the surfaces of molecules of neighbour fluid and parallel with surface of wave, vary their positions from the equilibrium state see Figure. (1). In general these variation of positions for the points on the each surface are equal to each other, function of two variables x (position) and t (time). We can show the variation with function $w(x, t)$. One can obtain relation between $w(x,t)$,  density and pressure of fluid. We consider a layer of vertical section S that is between the two parallel surfaces in positions $x$ and $x+dx$ in equilibrium state. The mass of this layer is $\rho_{0}Sdx$.    
Now consider this layer moves
to the  $w,\,w+\frac{\partial w}{\partial x}dx$ due
to right - pass wave (GWs in this work), see Figure 1. Then the
new volume of layer will vary to $Sdx\left(1+\frac{\partial w}{\partial x}\right)$.
The new volume generates a variation of the density while its mass
remains constant. Thus, by assuming the mass conservation, one writes
down 
\begin{equation}
\rho Sdx\left(1+\frac{\partial w}{\partial x}\right)=\rho_{0}Sdx.\label{eq: 34}
\end{equation}
Setting $s\equiv\left(\frac{\rho-\rho_{0}}{\rho_{0}}\right),$ the
above equation becomes 
\begin{equation}
(1+s)\left(1+\frac{\partial w}{\partial x}\right)=1.\label{eq: 35}
\end{equation}
As it is $s\approx\frac{\partial w}{\partial x}\approx10^{-4},$ the
product $s\frac{\partial w}{\partial x}$ can be neglect in Eq. (\ref{eq: 35})
and one gets 
\begin{equation}
s=-\frac{\partial w}{\partial x}.\label{eq: 36}
\end{equation}
For the sake of simplicity, one considers a perfect fluid with $P=P(\rho)$
during an adiabatic process. Then 
\begin{equation}
dP=\left(\frac{\partial P}{\partial\rho}\right)_{0}d\rho\quad with\quad d\rho=\rho-\rho_{0}.\label{eq: 37}
\end{equation}
 Setting $p\equiv dP$ and considering the speed of propagation of
the wave $\gamma$ given by $\gamma^{2}=\left(\frac{\partial P}{\partial\rho}\right)_{0},$
one gets 
\begin{equation}
p=\rho_{0}\gamma^{2}s.\label{eq: 38}
\end{equation}
Thus, combining Eqs. (\ref{eq: 36}) and (\ref{eq: 38}) one obtains
\begin{equation}
p=-\rho_{0}\gamma^{2}\frac{\partial w}{\partial x}.\label{eq: 39}
\end{equation}

\end{document}